\documentclass[aps,prb,groupedaddress,twocolumn,10pt]{revtex4-1}

\begin{document}

\newcommand{\be}{\begin{equation}}
\newcommand{\ee}{\end{equation}}
\newcommand{\bea}{\begin{eqnarray}}
\newcommand{\eea}{\end{eqnarray}}

\title{True SYK or (con)sequences}

\author{D. V. Khveshchenko}
\affiliation{Department of Physics and Astronomy, University of North Carolina, Chapel Hill, NC 27599}

\begin{abstract}
\noindent
Some generalizations of the Sachdev-Ye-Kitaev (SYK) model and different patterns of their reparametrization symmetry breaking are discussed. The analysis of such (pseudo)holographic
systems relates their generalized one-dimensional Schwarzian dynamics
to (quasi) two-dimensional Liouvillian quantum mechanics. As compared to the original SYK case, the latter might be dissipative or have discrete states in its spectrum, either of which properties alters thermodynamics and correlations while preserving the underlying $SL(2,R)$ symmetry.    
\noindent

\end{abstract}

\maketitle

\noindent
The $1+0$-dimensional (quantum mechanical)
SYK model of a large number $N$ of the Dirac (complex) \cite{sy}
of Majorana (real) \cite{kitaev} fermions with all-to-all random couplings and its various generalizations (including non-random ones) \cite{gurau} 
have attracted much attention lately. Those models are often cited as much-needed controllable examples of holographic correspondence which are expected to be dual to some  
$1+1$-dimensional gravity (plus, possibly, extra fields). 

Apart from remaining one of the central driving forces of modern string and high energy theory, the highly non-trivial and intriguing idea of (possibly, generalized beyond the original case of
 $AdS_5/CFT_4$) holography has already permeated other fields, including condensed matter physics. Over the past decade, a staggering number of holographic calculations alleged to be relevant to the 
realistic 'strange metals' and other complex ('non-Fermi-liquid') systems has emerged \cite{AdSCMT}. 

However, despite the
invariably upbeat claims of some of its enthusiastic practitioners, the use of the holographic technique 
outside of the original (subjected to a number of stringent constraints) string-theoretical context still remains to be justified and the true status of 
the cornucopia of look-alike and customarily verbose exercises in  
(generalized) classical relativity known under the acronym $AdS/CMT$ is yet to be ascertained.

In a sharp contrast with such 'per analogiam' (a.k.a. 'bottom up') approach,  
the SYK model seemed to offer an example of fully controllable holographic  
correspondence between two theories of different dimensionalities that 
might both be amenable to the (asymptotically) exact analytical treatments.
Also, despite its solubility, the SYK model was shown to be maximally chaotic, akin to black holes
\cite{kitaev,syk}, thus potentially providing insight into the inner workings 
of the generalized holographic conjecture and contributing towards its ultimate verification.  

It is known, though, that the (nominally $2d$) Jackiw-Teitelboim (JT) dilaton gravity that was conjectured as the SYK's bulk dual \cite{kitaev,syk} lacks any dynamical bulk degrees of freedom  
and, in fact, reduces to the theory of a fluctuating $1d$ boundary. Indeed, the only available 
solution with constant (negative) curvature $R=-2$ implies the rigid $AdS_2$ metric.  
Correspondingly, the JT spectrum turns out to be quite different from 
the 'dimension gap-free' SYK's one and in order to reconcile between the two 
an infinite tower of additional massive bulk scalar fields needs to be introduced \cite{syk}.

Thus, while important and insightful, 
the SYK-JT relationship may not quite rise to the same level as a would-be 'bona fide'  holographic correspondence between two theories operating irreducibly 
(yet, in both cases, locally) in different dimensions \cite{AdSCMT}.  
Instead, the SYK-JT correspondence where the gravitational background metric appears to be 
essentially non-dynamical could be viewed as a case of 'holography light'  - a less ambitious scenario \cite{dvk1} that so far has been largely ignored.    

As a matter of fact, nearly all of the previous 'bottom up' holographic 
calculations \cite{AdSCMT} have been performed for certain fixed (usually, well-studied)
background metrics while neglecting any potential $O(1/N)$ (here $N$ stands for the rank of an internal symmetry group, if any) corrections - either matter's backreaction upon gravity or quantum fluctuations of the latter. 

Notably, though, the thus-obtained results would then 
be used to seek (and often claimed to have found) a good (including quantitative) 
agreement with some pre-selected sets of data on the physical systems with $N\sim 1$.   
Conceivably, if indeed present, such a fortuitous agreement would seem to indicate that 
no metric fluctuations should have been allowed in the first place.   

Apart from the continuing exploration of such salient features of the original SYK model as its maximally chaotic behavior \cite{kitaev,syk}, some of the recent work has been challenging the various popular simplifying assumptions, such as replica symmetry of the SYK solutions \cite{replica}. Also, a technically related issue of the possibility of spatially dispersive solutions has been raised \cite{dvk2} in the context of the various multi-dimensional 'SYK-lattice' generalizations which would be routinely 
assumed to remain spatially ultra-local due to the (presumed 
to always remain intact) local $Z_2$ symmetry \cite{lattice}. 

The work on the SYK generalizations targets those behaviors that could survive departures 
from the original model, thus manifesting potentially generic, rather than unique to the SYK model, properties. In the present note, this quest is further pursued towards the  
different patterns of such central to the SYK issue as reparametrization symmetry breaking and
the associated sequence of SYK-like models.

By analogy with the original SYK treatment \cite{sy,kitaev,syk}
a convenient starting point can be chosen in the form of 
a $1+0$-dimensional path integral over a pair of 
bi-local field variables $G(\tau_1,\tau_2)$ and $\Sigma(\tau_1,\tau_2)$ 
\bea 
Z=\int D{G}D{\Sigma} 
({\it Det}[F[\partial_{\tau}]+\Sigma])^N
~~~~\\
\exp(N\int_{\tau_1,\tau_2} {G}{\Sigma}-A[{G}]))\nonumber
\eea
where $F$ and $A$ are some functionals of their (operator-valued) arguments
and the thermodynamics time varies within the interval $0\leq\tau\leq 1/\beta$.
Depending on the nature of the underlying fundamental fields - e.g., complex vs real
fermions - the variables $G$ and $\Sigma$ may have certain symmetrical  
properties as well. In what follows they will be treated as 
real-valued, consistent with the Majorana case. 
A generic functional 
\be
A=N\sum_k^{\infty}\int_{\tau_1,\dots\tau_k}
J^2_k(\tau_1,\dots\tau_k)
G^q(\tau_1,\tau_2)\dots G^q(\tau_{k-1},\tau_k)
\ee
reproduces the '$2q$-interacting' SYK model for $J^2_k\sim \delta_{k2}$  
(in contrast to its original 
formulation \cite{sy,kitaev}, here $q$ is an arbitrary - not necessarily even - integer).
In fact, the action (2) is not unique to the original
SYK model but can also describe its non-random cousins \cite{gurau}. 

From the general standpoint, 
its applicability of Eq.(2) requires asymptotic dominance of the 'chain-melonic' diagrams. For a prospective microscopic system such behavior (or a lack thereof)
could, in principle, be established by extending the analysis \cite{melon} of the 
SYK diagrammatics which is dominated by the ordinary 'melonic' ($k=2$) graphs.  
 The standard renormalization argument suggests, however, that in the absence of
a special fine tuning of the parameters $J_k$ the infrared (IR) 
dynamics of the theory (1,2) is still likely to
be governed by the lowest dimension term (i.e., $k=2$). 

Varying Eq.(1) with respect to $G$ and $\Sigma$ one obtains the mean-field equations  
\bea
\int_{\tau}(F(\partial_{\tau})\delta(\tau_1,\tau)+\Sigma(\tau_1,\tau))G(\tau,\tau_2)=
\delta(\tau_1-\tau_2)
\nonumber\\
\Sigma(\tau_1,\tau_2)={1\over N}{\delta A\over \delta G(\tau_1,\tau_2)}~~~~~~~~~~
\eea
Neglecting the time derivatives completely, one finds 
that the saddle points of (1) are given by the solutions of the integral equation
\be 
\int_{\tau}G(\tau_1,\tau){\delta A\over \delta G(\tau,\tau_2)}=\delta(\tau_1-\tau_2)
\ee
which appears to be manifestly invariant under the infinite group $Diff(S^1)$ of reparametrizations of the thermal circle $\tau\to f(\tau)$ with the periodicity condition 
$f(\tau+\beta)=f(\tau)+\beta$, provided that $G$ and $\Sigma$ transform as
\bea 
G(\tau_1,\tau_2)\to G_f=[f^{\prime}(\tau_1)f^{\prime}(\tau_2)]^\Delta G(f(\tau_1),f(\tau_2))\nonumber\\
\Sigma(\tau_1,\tau_2)\to \Sigma_f=[f^{\prime}(\tau_1)f^{\prime}(\tau_2)]^{1-\Delta}\Sigma
(f(\tau_1),f(\tau_2))
\eea
where $\Delta=1/2q$.

In the SYK case Eq.(4) permits a translationally-invariant 'conformal' solution
(here $\delta\tau_{12}=\tau_1-\tau_2$) \cite{sy,kitaev}
\be 
G_0(\tau_1,\tau_2)=({\pi\over \beta\sin(\pi\delta\tau_{12}/\beta)})^{2\Delta}  
\ee
which spontaneously breaks the full reparametrization symmetry down to its 
three-dimensional subgroup $SL(2,R)$  implemented through the Mobius transformations 
\be 
\tan {\pi f(\tau)\over \beta}\to{a\tan{\pi f(\tau)\over \beta}+b\over c\tan {\pi f(\tau)\over \beta}+d}
\ee
with $ad-bc=1$, under which the function (6) (hence, the entire action given by Eqs.(1,2)) remains 
invariant. 

The rest of the group $Diff(S^1)/SL(2,R)$ then extends Eq.(6) onto the coadjoint Virasoro orbit where the dynamics of the field variable $f(\tau)$ is  
governed by the (subdominant) non-reparametrization invariant action (1,2). 

In the original SYK model ($F(x)=x, J_k=\delta_{k2}$) 
the leading (albeit IR-irrelevant in the RG sense)
term of order $O(N/\beta J)$ stems from the first time derivative in the (Pfaffian) determinant   
\be 
A_0=Tr\ln(1-\partial_{\tau}G_f)=-M
\int_{\tau} Sch{\{}\tan{\pi f\over \beta},\tau {\}}
\ee
and is controlled by the characteristic time scale
$M={\alpha_s N/J}$ proportional to the numerically 
computed ($q$-dependent) prefactor $\alpha_s$ \cite{kitaev,syk}. The  
integrand readily identifies with the manifestly geometrical and $SL(2,R)$-invariant Schwarzian derivative
\be 
Sch {\{} F,\tau {\}}={F^{\prime\prime\prime}\over F^{\prime}}-{3\over 2}
({F^{\prime\prime}\over F^{\prime}})^2 
\ee
which satisfies the differential 'chain rule',
$
Sch {\{} F(f),\tau {\}}=Sch {\{} F,f {\}}{f^\prime}^2 
+Sch {\{} f,\tau {\}}
$, 
when applied to a composite function, such as $F(f(\tau))=\tan\pi f(\tau)/\beta$
(alternatively, the thermal circle can also be
parametrized in terms of the function $e^{i\pi f/\beta}$)  \cite{kitaev}.

The emergence of the Schwarzian is to be expected as it
controls the short-time expansion of the transformed solution (6) 
(here $\tau=(\tau_1+\tau_2)/2$)
\bea 
\delta G=G_f(\tau_1,\tau_2)-G_0(\tau_1,\tau_2)\approx\\ 
\approx 
{\Delta\over 6}(\delta\tau_{12})^2 
Sch {\{} f,\tau {\}}G_0(\tau_1, \tau_2)+\dots\nonumber
\eea
The standard mean-field ('large-$N$') SYK scenario \cite{sy,kitaev} 
sets in for $1/J\ll\beta\ll M$ where 
the fluctuations $\delta G$ about the solution (6) are negligible. 
In contrast, for $M\lesssim\beta$ these fluctuations grow strong, 
thereby modifying the mean-field behavior \cite{bak}.  

Notably, the next,
$O(N/(\beta J)^2)$, order correction to Eq.(8) is no longer local \cite{kitaev}
\be
{\delta A}\sim {N\over J^2}
\int_{\tau_1\tau_2}
{(f^{\prime}_1f^{\prime}_2)^2
\over (\delta\tau_{12})^4}\ln({J^2
(\delta\tau_{12})^2\over f^{\prime}_1f^{\prime}_2}) 
\ee 
so the holographic postulate of locality holds only in the leading approximation. 

Moreover, in addition to the 'gradient' terms, such as Eq.(8), 
the reparametrization symmetry can also be broken 
by choosing more generic time-dependent couplings $J_k$ in Eq.(2). Furthermore, for certain couplings 
the $SL(2,R)$-invariant Eq.(6) still remains a solution as, e.g., in the case of  
\be 
J^2_{k}(\delta\tau)=\delta_{k,2}{J^{2-2\gamma}\over (\delta\tau)^{2\gamma}}
\ee
although $\gamma>0$ alters the anomalous field dimension to
\be 
\Delta={1-\gamma\over 2q}
\ee
The low-energy soft-mode action then 
gets modified by a non-local term which for $\gamma\ll 1$ can be approximated as a quadratic one
(here $\Gamma=2q\gamma NJ\Delta/3$)
\bea
{\delta A}={\Gamma\over J}
\int_{\tau_1\tau_2}(\delta\tau_{12})^2\ln (J{\delta\tau_{12}}) 
G^{2q}_f(\tau_1,\tau_2)Sch{\{}\tan{\pi f\over \beta},\tau {\}}\approx\nonumber\\
\approx {\Gamma\over J}\int_{\tau_1\tau_2}
{(f^{\prime}_1-1)(f^{\prime}_2-1)\over (\delta\tau_{12})^2}~~~~~~~~
\eea
and is reminiscent of the Ohmic dissipation in the Caldeira-Leggett model. 

Among other things, the underlying $SL(2,R)$ algebraic structure 
suggests a systematic way of extending the conjectured 
(pseudo)holographic SYK-JT connection 
from the pure Schwarzian (8) to a broader class of the $1d$ boundary 
theories whose action may include Eqs.(11),(14), etc. 
They can be conveniently formulated in terms of the Hamiltonian dynamics on the 
$4d$ phase space spanned by two pairs of canonically conjugated variables, $(f,\pi_f)$ and $(\phi,\pi_{\phi})$, the former one being the aforementioned boundary 'conformal' time \cite{verlinde}.  

A pertinent Hamiltonian then conforms to the $SL(2,R)$ -invariant quadratic Casimir operator
\be 
H={1\over 2}L^2_0-{1\over 4} (L_1L_{-1}+L_{-1}L_1)
\ee
where the $SL(2,R)$ generators $L_{0,\pm 1}$ obey the Poisson brackets algebra   
\be 
{\{} L_0,L_{\pm 1} {\}}=\pm L_{\pm 1},~~~~~~~ {\{} L_{-1},L_{1} {\}}=2L_{0}
\ee
The various realizations of this algebra allow one to construct a host 
of dual boundary systems.  

For instance, the general Hamiltonian (15) constructed with the use of the ansatz      
\bea
L_{-1}=\pi_f,~~~~~~~L_0=f\pi_f+\pi_{\phi},~~~~~~~~\\
L_1=f^2\pi_f+2f\pi_{\phi}+A(\phi)-B(\phi)\pi_f-{C(\phi)\over \pi_f}\nonumber
\eea
describes a charged non-relativistic particle of unit mass 
confined to a $2d$ surface with some (diagonal) metric 
$g_{ij}=diag [g_{\phi\phi}, g_{ff}]$ and 
subjected to vector $(A_{\phi},A_f)$ and scalar 
$\Phi$ potentials
\be  
H={1\over 2}g^{\phi\phi}{\pi^2_{\phi}}+{1\over 2}g^{ff}(\pi_f-A_f)^2+\Phi
\ee
where the background fields are given by the expressions 
\bea   
g^{ij}=diag[1,B(\phi)],~~~\\
A_i=(0,{A(\phi)\over B(\phi)}),~~~\Phi=C(\phi)-{A(\phi)^2\over 4B(\phi)}\nonumber
\eea
The coordinate $f$ appears to be cyclic, so that the conjugate momentum $\pi_f$ is conserved. In fact, even for generic vector $A_i(\phi)$ and scalar $\Phi(\phi)$ potentials  
the dynamics described by the Hamiltonian (18)
remains effectively one-dimensional (this observation would have been far less obvious, though, had the vector potential been taken in a gauge other than the Landau one).

Moreover, for $A(\phi)=2ae^{\phi}$,  $B(\phi)=be^{2\phi}$, and $C(\phi)=c$ 
the operators (17) obey the algebra (16), thus guaranteeing the $SL(2,R)$-invariance of Eq.(15)
which now takes the form 
\be  
H={1\over 2}{\pi^2_{\phi}}+{b\over 2}\pi^2_f e^{2\phi}-ae^{\phi}\pi_f+{1\over 2}c
\ee
while the metric $ds^2=d\phi^2+e^{-2\phi}df^2$ becomes that of the hyperbolic plane $H^2$.

Such connection between the SYK problem and a particle on $H^2$ in magnetic field has been pointed out and exploited before \cite{kitaev,verlinde}. 
As the above suggests, it can be extended towards a broader class of (ostensibly) $2d$ Hamiltonians - albeit, at the expense of adding new (admittedly, somewhat unphysical) 
terms proportional to the powers of momentum $\pi^n_f$ with $n>2$ and/or $n<0$.  

The standard SYK scenario corresponds to choosing 
$a\pi_f=-\mu, b=c=0$ which reduces (20) to 
the Hamiltonian of the $1d$ Liouville quantum mechanics  
\be  
H={1\over 2}\pi^2_{\phi}+\mu e^{\phi}
\ee 
whose relation to the Schwarzian action (8) has been discussed extensively \cite{bak,verlinde}. 
Indeed, upon the substitution 
$ 
f^\prime=e^{\phi}
$
the zero-temperature action (8) amounts to 
the Gaussian kinetic energy of the (unbounded) variable $\phi(\tau)$ since 
$Sch{\{} f,\tau {\}}=({\partial_{\tau}\phi})^2$.
 
Moreover, this change of variables can be formally implemented as a constraint enforced by
the momentum $\pi_{f}$ playing the role of the Lagrange multiplier \cite{bak,verlinde}.
Introducing the second momentum $\pi_{\phi}$ in the 
Legendre transformation of the Schwarzian and rescaling the entire action with $M$    
one then recovers the $2d$ Lagrangian 
\be 
L=\pi_{\phi}{\phi^{\prime}}-{1\over 2M}\pi_{\phi}^2+
\pi_f({f^{\prime}}-e^{\phi})
\ee
which implies a constant $\pi_f=\mu\sim J$ \cite{bak}, consistent with (21). 
At finite temperatures the above change of variables results in the additional term 
$\delta L=-(2\pi/\beta)^2e^{2\phi}$ in (22) \cite{verlinde}.
By contrast, in Ref.\cite{bak} a modified finite-temperature relation between $f$ and $\phi$ was used, $(\tan\pi f/\beta)^{\prime}=e^{\phi}$, which yields the same $1d$ Liouville Hamiltonian (21) at all temperatures. However, in this case the variable $f$ ceases to be 
cyclic which makes the intrinsically $1d$ nature of this theory more obscure.  

Quantizing the 'particle-in-magnetic-field' 
Hamiltonian (20) for $a={\cal A}/M, b=1/M$, and $c={\cal A}^2/4M$,
factorizing its eigenstates, $\Psi(\phi,f)=\psi(\phi)e^{i\mu f}$,
and shifting the variable $\phi\to\phi-\ln {\cal A}/\mu$,  
one arrives at the $1d$ Schroedinger equation   
\be
(-{\partial^2\over \partial{\phi}^2}+
\lambda^2(e^{2\phi} - 2 e^{\phi}sgn\mu ))\psi=(\epsilon-\lambda^2)\psi
\ee
where $\epsilon=2M(E-c)$ and $\lambda=a/2b^{1/2}$
(in the finite-temperature case 
$\lambda=\mu\beta=O(\beta J)\gg 1$). 

In the previous studies of the SYK model the sign of $\mu=\pi_f$ would 
be routinely chosen negative and the squared exponential 
term $e^{2\phi}$ neglected 
(alternatively, in Ref.\cite{bak} the latter would have never appeared in the first place) 
so as to reproduce the monotonic repulsive potential of the Liouville 
Hamiltonian 
anticipated on the basis of the correspondence with the bulk Euclidean $AdS_2$ \cite{kitaev,syk}. 

For $\mu<0$ the positive definite  spectrum of (23) is 
continuous, $\epsilon_k=(k^2+1/4+\lambda^2)$, 
parametrized by a 'momentum' $k$, while its eigenstates are 
given by the Whittaker function 
(here $z=2\lambda e^{\phi}$) 
\be
\psi_{k}\sim e^{-\phi/2} W_{\lambda, ik}(z)
\ee
For $b=0$ (24) reduces to the eigenstates of (21) given by 
the modified Bessel functions, $\psi_k\sim K_{2ik}({\sqrt z})$ \cite{bak,verlinde}.

In terms of the eigenstates $\psi_k$ the partition function
given by the (non-Gaussian) path integral can be computed as 
\bea 
Z(\beta)=\int^{\phi(\beta/2)=\phi_0}_{\phi(-\beta/2)=\phi_0}D\phi
e^{-\int_\tau L(\phi)}=\nonumber\\
=\int^{\infty}_0dk|\psi_k(\phi_0)|^2e^{-E_k\beta}
\eea
For $b=0$ this calculation yields 
the free energy of the SYK model \cite{kitaev,syk,bak}
\be 
F=-{1\over \beta}\ln Z(\beta)=E_0-{S_0\over \beta}-{2\pi^2M\over \beta^2}
+{\pi^2\mu N\over 6\beta^3J^2}+{3\over 2\beta}\ln\beta J
\ee
where $E_0$ and $S_0$ are the extensive 
ground state energy and zero-temperature ('residual') entropy.
The last two terms represent the next order corrections ($O(1/J\beta)$ and $O(1/N)$, 
respectively) \cite{kitaev,syk}.

Using (25) one finds the (many-body) density of states (DOS) 
\be 
\rho(\epsilon)={1\over 2\pi i}\int_{\beta}e^{\beta E}Z(\beta)\sim
e^{S_0}\sinh(2\pi{\sqrt \epsilon})
\ee
Alternatively, this result can be inferred from the exact DOS of the 'particle-in-magnetic-field' problem \cite{comtet}, 
$
\rho(\epsilon)\sim\sinh 2\pi{\sqrt \epsilon}/(cosh 2\pi{\sqrt \epsilon}+\cos 2\pi\lambda)
$,
by shifting the field $\phi\to\phi-ln(-2\lambda)$ and taking the limit $\lambda\to i\infty$
 \cite{kitaev,verlinde}.

In contrast to the Liouville scenario, for $sgn \mu =1$ Eq.(23) features 
the non-monotonic Morse potential and may possess additional discrete states  given by the associated Laguerre polynomials
\be
\psi_{n}(z)\sim
z^{\lambda-n-1/2-z/2} L_n^{2\lambda-2n-1}(z)
\ee
at the discrete energies  
$\epsilon_n=-(n-\lambda+1/2)^2$, $n=0,\dots, [\lambda-1/2]$.

At low temperatures ($\mu\beta\gg 1$) the number 
$[\lambda-1/2]$ of the bound states is large   
and they dominate the partition function. 
Moreover, their spectrum becomes almost equidistant, allowing one 
to replace the actual Morse potential with the approximate quadratic ('oscillator') one.

Furthermore, in the presence of the reparametrization symmetry-breaking term (14)
the effective action becomes that of a 'damped Morse potential'.
Although canonical quantization of a dissipative system can be intrinsically problematic, 
one can still resort to the path integral approach
to study its statistical mechanics and correlations.

In the quadratic approximation, one then obtains the Gaussian action  
(here $\Omega\sim \mu\beta/M$)
\be
\delta S={M\over 2}
\sum_{n}
(\omega^2_n+\Omega^2+{\Gamma}|\omega_n|) |\phi_n|^2
\ee  
Although the quadratic action (29) is gapped, the dynamics of the 
conformal time $f$ still features the zero 
modes $n=0, \pm 1$ as the deformed Schwarzian action given by Eqs.(8,14) 
remains invariant under the $SL(2,R)$ group
(this can also be inferred from the relation $\phi\approx f^\prime-1$). 

Using the simplified Eq.(29) one can compute the partition function 
\be 
Z(\beta)
={1\over 2\sinh\beta\Omega/2}
\prod^{\infty}_{n=1}{\Omega^2+\omega_n^2\over \Omega^2+\omega_n^2+\Gamma|\omega_n|}
\ee
where the divergent product can be regularized by introducing the cutoff 
frequency $\omega_{max}\sim J$.

This way one obtains the free energy 
\bea 
{F\over N}={1\over \beta}\ln(\beta\Omega)+{1\over \beta}\sum^{\omega_{max}}_{n=1}\ln(1+{\Gamma|\omega|\over {\omega_n^2+\Omega^2}})\approx\nonumber\\
\approx {\Omega\over 2}+{1\over \beta}\ln(1-e^{-\beta\Omega})+{\Gamma\over 2\pi}\ln({J\over \Omega})
\eea
for $1/\beta \ll \Omega, \Gamma$, whereas at higher temperatures the quadratic approximation fails
and the leading part of Eq.(26) would be reproduced instead. 
The thermodynamics properties of the Morse model are, therefore, markedly different from those of the Lioville one. In particular, the specific heat is exponentially suppressed.

From (31) one infers the oscillator-like "Dirac comb' DOS
which averages out to a constant at
energies $E\gg\Omega$
\be 
\rho(\epsilon)\approx{\lambda-{1\over 2}\over 2\pi}
\sum_n\delta(E_n-\Omega(n+{1\over 2}))\sim M
\ee
For a more detailed comparison with the standard SYK case 
one can also evaluate the correlator of the stress tensor
$T(\tau)=M(f^{\prime\prime\prime}-(2\pi/\beta)^2f^{\prime})$ 
\bea
<T(\tau)T(0)>=
M\sum_{n}{e^{2\pi\tau/\beta}(\omega_n^2-(2\pi/\beta)^2)\omega^2_n
\over {\omega^2_n+\Omega^2+\Gamma|\omega_n|}}\nonumber\\
\sim M max[1/\beta^3,\Omega^3]\sin\Omega\tau e^{-\Gamma\tau/2}~~~~~~~~
\eea  
At $\tau=0$ this result agrees with the direct estimation of the 
energy variance (for $\Gamma<\Omega$)
\be
<(\delta E)^2>={\partial^2\over \partial\beta^2}\ln Z(\beta)\sim M max[1/\beta^3,\Omega^3]
\ee 
As a more subtle diagnostic of the boundary dynamics, 
in the 'Schwarzian' (long-time, low-temperature, $M\lesssim\beta$) limit the fluctuations of the Liouville soft mode strongly affect the averaged products  
\bea 
<G_f(\tau_1,\tau_2)\dots G_f(\tau_{2p-1},\tau_{2p})>=\nonumber\\
=\int D\phi \prod_{i=1}^p
{e^{\Delta(\phi(\tau_{2i-1})+\phi(\tau_{2i}))}\over (\int_{\tau_{2i-1}}^{\tau_{2i}}e^{\phi})^{2\Delta} }
e^{-\int_{\tau}L(\phi)}
\eea
The denominator  in (35) can be promoted to the exponent 
where it contributes to the overall piece-wise (in the time domain) Liouville potential 
representing the $2p$ consecutive quenches \cite{bak}. 

As the result, in the original SYK case with $q=2$ 
the fluctuation-dressed averages $<G^p_f(\tau,0)>$ were found to 
change their $p$-dependent algebraic decay $\sim 1/\tau^{p/2}$ for $\tau\ll M$ 
to the universal behavior $\sim 1/\tau^{3/2}$ developing for for $\tau\gg M$  \cite{bak}. 

By contrast, in the Morse theory for $\Omega\ll 1/M$
the fluctuation-averaged two-point correlator changes its behavior from (6) for $\tau\ll M$
to
\bea
<G_f(\tau_1,\tau_2)>\approx\sum_{n}e^{-E_n\tau}N_1(E_n)\sim 1/\tau,~~~{M}\ll\tau\ll{1\over \Omega}\nonumber\\
\sim e^{-\Omega\tau},~~~\tau\gg{1\over \Omega}~~~~~~~
\eea 
since the matrix element $N_1=|<0|e^{\Delta\phi}|n>|^2$ and the DOS (32) are non-singular for $n\le\lesssim\lambda$ and the sum over $\sim\lambda$ terms 
can be approximated by the integral. In the opposite limit, $1/M\lesssim \Omega$, 
there is no room for algebraic behavior and (36) decays exponentially. 

Likewise, the SYK averages of the higher powers $<G^p_f(\tau,0)>$ are sensitive to the
behavior of the $p$-particle DOS and under the above conditions 
demonstrate the crossovers  
from the short-time power-law behavior $\sim 1/\tau^{p/q}$ to the intermediate
universal one, $\sim 1/\tau$, and, finally, to the exponential decay $\sim e^{-\Omega\tau}$
at the longest times. In the latter limit, the would-be universal algebraic
contribution of the continuous part of the spectrum, $1/\tau^{6\Delta}$,
is suppressed by the much smaller factor $\sim e^{-\Omega\lambda\tau}$.

In addition to 
the Schwarzian fluctuations, the multi-point correlators can also receive subleading  
$O(1/N)$ contributions from the massive modes which are not governed 
by the Schwarzian action (8) but can be accounted for 
by summing the ladder diagrams \cite{kitaev,syk}.  

Specifically, in the case of the $p=2$ function $<G_f(\tau_1,\tau_2)G_f(\tau_3,\tau_4)>$ 
the massive modes do not significantly contribute  for $\tau_{2,3}<\tau_{1,4}$,  
whereas in the domain $\tau_{1,2}<\tau_{3,4}$ they do, being solely responsible for the irreducible (non-factorizable) contribution to that function \cite{bak}. 

In particular,  
upon analytically continuing from the domain $\tau_4<\tau_2<\tau_{3}<\tau_{1}$
to the real times $\tau_{1}=\beta/4-it/2, \tau_2=-\beta/4-it/2,
\tau_3=it/2$,  $\tau_{4}=-\beta/2+it/2$,  the all-important out-of-time-order (OTO) correlators  
demonstrate their initial short-time/high temperature exponential growth 
\be
{<G_f(\tau_1,\tau_3)G_f(\tau_2,\tau_4)>\over <G_f(\beta/2,0)>^2}= 1-O({\beta\over M})e^{\lambda_Lt}
\ee
revealed by summing the 'causal' ladder series for $\beta\ll t\ll\beta\ln M$ \cite{kitaev,syk}. 

In the Morse case the counterpart of (37) exhibits the less-than-maximal Lyapunov exponent
\be 
\lambda_L={2\pi\over \beta}(1-O(\gamma))
\ee
which should be contrasted with the result obtained in the original SYK model
$\lambda_L=2\pi/\beta(1-O(1/\beta J))$ \cite{kitaev,syk}.

Also, in the intermediate-time regime (present for $\Omega\ll 1/M$) 
the Morse OTO function decays with the real time $t$ as 
\be 
<G_f(\tau_1,\tau_2)G_f(\tau_3,\tau_4)>\sim 1/t^4
\ee
which dependence again differs from the $\sim 1/t^6$ asymptotic found 
in the $q=2$ SYK model \cite{bak}. 

Returning to the general question of the holographic principle's 
implementation  in the SYK-like models, 
one finds that some intrinsically $1d$ details of the Scharzian/Liouville theories (or, more generally, the 'particle-in-magnetic-field' problem) 
can also be observed in their conjectured bulk duals.

Specifically, in the  
JT theory the same $1d$ differential equation (23) emerges, its spectrum now being that of the $SL(2,R)$ Casimir operator formulated in terms of two angular variables, 
$\theta$ and $\varphi$, related to the space-time coordinates 
($\tan(\theta-\varphi/2)=re^{-\tau}$, $\tan(\theta+\varphi/2)=re^{\tau}$). The wave function once again factorizes onto 
the 'angular' and 'radial' ones, $e^{im\varphi}\chi_k(\theta)$, 
the latter obeying the $1d$ equation \cite{kitaev,syk} 
\be
(-{\partial^2\over \partial{\theta}^2}-{k^2+{1\over 4}\over \cos^2\theta}+2\lambda m\tan\theta)
\chi=(m^2-\lambda^2)\chi
\ee
solved by the functions (24). 

Alternatively, the tangle of (preudo)holographic 
relationships between the $SL(2,R)$-symmetric boundary (Schwarzian/Liouville-like) and bulk (JT-like) theories can be viewed as different forms of embedding (at fixed radial and angular vs temporal and angular coordinates, respectively) into the global $AdS_3$ space 
\cite{verlinde,AdS3}. 

To further elucidate such generalized relationship one can use the framework of the generalized 
JT-dilaton theory 
\be
S=\int_{\tau,r}(R\Phi+U(\Phi)){\sqrt g}+\int_\tau K\Phi
\ee
where the appropriate dilaton potential $U(\Phi)$ might be able to reflect  
the various $Diff(S^1)$-symmetry breaking extensions to the basic Schwarzian action.

To that end, the quadratic term $U_2(\Phi)\sim\Phi^2$  
has already been shown to correspond to the non-local contribution (11) \cite{kitaev}. 
In the future, it would be interesting to establish a link between the 
higher order terms $\Phi^n$ and (non)local and/or (non)geometric 
deformations of the fundamental Schwarzian action (8).  

In the meantime, the natural emergence of the 
'particle-in-magnetic-field' (Liouville-like) quantum mechanics 
provides a convenient technical framework for generating 
novel sequences of (pseudo)holographic $1d$ systems and their (ostensibly) $2d$ duals.
The latter would then be essentially topological, akin to the theory 
of incompressible electron droplets in the Quantum Hall Effect \cite{QHE}.  
 
Still more examples of such 
correspondence can be discovered by studying other Virasoro orbits  
as well as the even more general problem of quantum mechanics on 
a wider class of group manifolds and their cosets \cite{coset}. 

To conclude, the systematic (non-conformal) extensions of the SYK model can be achieved along the lines of single-particle quantum mechanics, thus allowing one to further explore their putative 
bulk (albeit, dimensionally reducible) duals. Continuing such work may help to elucidate the true status of the holographic studies of the concrete (first of all, condensed matter) systems.

\end{document}